\documentclass[conference]{IEEEtran}
\IEEEoverridecommandlockouts
\usepackage{cite}
\usepackage{amsmath,amssymb,amsfonts}
\usepackage{algorithmic}
\usepackage{graphicx}
\usepackage{textcomp}
\usepackage{xcolor}
\def\BibTeX{{\rm B\kern-.05em{\sc i\kern-.025em b}\kern-.08em
    T\kern-.1667em\lower.7ex\hbox{E}\kern-.125emX}}
\begin{document}

\title{Cybersecurity and Frequent Cyber Attacks on IoT Devices in Healthcare: Issues and Solutions\\

}

\author{\IEEEauthorblockN{Zag ElSayed}
\IEEEauthorblockA{\textit{School of Information Technology} \\
\textit{CECH, University of Cincinnati}\\
Ohio, USA \\
}
\and
\IEEEauthorblockN{Ahmed Abdelgawad}
\IEEEauthorblockA{\textit{School of Eng. \& Tech.} \\
\textit{Central Michigan University}\\
Michigan, USA\\
}
\and
\IEEEauthorblockN{Nelly Elsayed}
\IEEEauthorblockA{\textit{School of Information Technology} \\
\textit{CECH, University of Cincinnati}\\
Ohio, USA \\
}
}

\maketitle

\begin{abstract}
Integrating Internet of Things (IoT) devices in healthcare has revolutionized patient care, offering improved monitoring, diagnostics, and treatment. However, the proliferation of these devices has also introduced significant cybersecurity challenges. This paper reviews the current landscape of cybersecurity threats targeting IoT devices in healthcare, discusses the underlying issues contributing to these vulnerabilities, and explores potential solutions. Additionally, this study offers solutions and suggestions for researchers, agencies, and security specialists to overcome these IoT in healthcare cybersecurity vulnerabilities. A comprehensive literature survey highlights the nature and frequency of cyber attacks, their impact on healthcare systems, and emerging strategies to mitigate these risks.
\end{abstract}


\section{Introduction}
The Internet of Things (IoT) has seen widespread adoption in the healthcare sector \cite{HAGHIKASHANI2021103164}, facilitating real-time patient monitoring, remote diagnostics, and efficient healthcare delivery \cite{GOVIND2021106933}. IoT technology facilitates the creation of an interconnected ecosystem where devices, sensors, and systems seamlessly exchange data, enabling real-time monitoring, enhanced diagnostic capabilities, and personalized patient care \cite{beke2019trends}. In healthcare, IoT encompasses a broad spectrum of devices such as wearable health monitors, implantable devices, innovative medical equipment, and sophisticated healthcare information systems. These innovations collectively contribute to improved patient outcomes, operational efficiency, environment saving, and cost-effectiveness \cite{saleem2020iot}. 

\begin{figure}[htbp]
\centerline{\includegraphics[width =\linewidth]{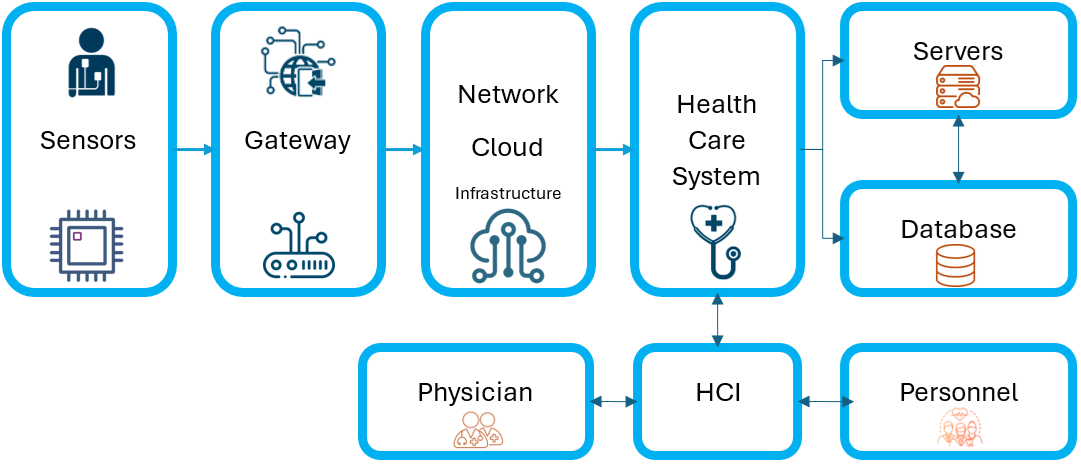}}
\caption{The current popular IoT infrastructure architecture.}
\label{fig1}
\end{figure}

However, integrating IoT devices into healthcare infrastructure is not without its challenges. The attributes that make IoT devices valuable—ubiquity, connectivity, and data richness—also render them susceptible to a myriad of cybersecurity threats \cite{8399126}, a traditional Healthcare IoT infrastructure shown in Fig.~\ref{fig1}. As healthcare systems increasingly rely on IoT devices to manage critical medical data and deliver essential services, the potential for cyber attacks escalates. These attacks can compromise patient safety, disrupt healthcare services, and damage financial and reputation.

The complexity of IoT environments in healthcare systems arises from the heterogeneity of devices, diverse communication protocols, and varying security standards. Each device within the IoT ecosystem represents a potential point of vulnerability that malicious actors can exploit. Common attack vectors include malware, ransomware, denial-of-service (DoS) attacks, man-in-the-middle (MitM) attacks, and phishing schemes. These attacks can lead to unauthorized access to sensitive patient data, manipulation of medical devices, and interruption of healthcare services.

Addressing the cybersecurity challenges associated with IoT in healthcare requires a multi-faceted approach that involves enhancing device security, implementing robust network security measures, adhering to regulatory compliance, and fostering a culture of security awareness among healthcare professionals. Among the key strategies to mitigate these risks are encryption, strong authentication mechanisms, regular firmware updates, network segmentation, and intrusion detection systems.

In addition to technical measures, it is imperative to consider the regulatory landscape governing healthcare IoT devices. Regulations such as the Health Insurance Portability and Accountability Act (HIPAA) in the United States and the General Data Protection Regulation (GDPR) in the European Union set stringent data protection and privacy standards. Compliance with these regulations not only ensures legal adherence but also reinforces the security posture of healthcare organizations.

This paper aims to provide a comprehensive review of the cybersecurity threats facing IoT devices in healthcare, examine the contributing factors to these vulnerabilities, and explore potential solutions and strategies for mitigating these risks. By synthesizing current research and offering practical recommendations, this study aims to assist researchers, agencies, and security specialists in fortifying the security of healthcare IoT systems and safeguarding patient data.

\section{Overview of IoT in Healthcare}
The deployment of IoT devices in healthcare represents a significant technological advancement that has the potential to revolutionize patient care, diagnostics, and treatment. These devices facilitate continuous health monitoring, real-time data analysis, and personalized medical interventions. The various types of IoT devices are commonly used in the healthcare sector, and many classifications for IoT Healthcare devices are mentioned in the literature. For example, in~\cite{pradhan2021iot} classified the IoT devices into Identification, Location, and communication technologies. In~\cite{hameed2020intelligent}, the authors classified the device into sensors, Smart Computing, Security, Remote monitoring, big data, connectivity, and lab or chip. However, that classification is missing the security vision of the architecture.

\begin{figure}[htbp]
\centerline{\includegraphics[width =\linewidth]{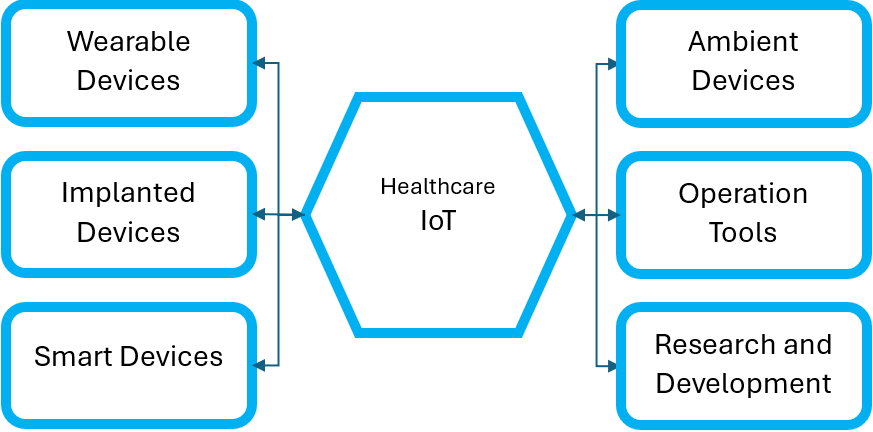}}
\caption{Proposed Threat oriented Healthcare IoT devices classification.}
\label{fig2}
\end{figure}

Thus, we came up with a more security and threat-oriented classification, which includes wearable devices, implanted devices, smart medical devices, ambient devices, research and development, and healthcare operation systems. These classes are shown in Fig.\ref{fig2}.

\begin{figure}[htbp]
\centerline{\includegraphics[width =\linewidth]{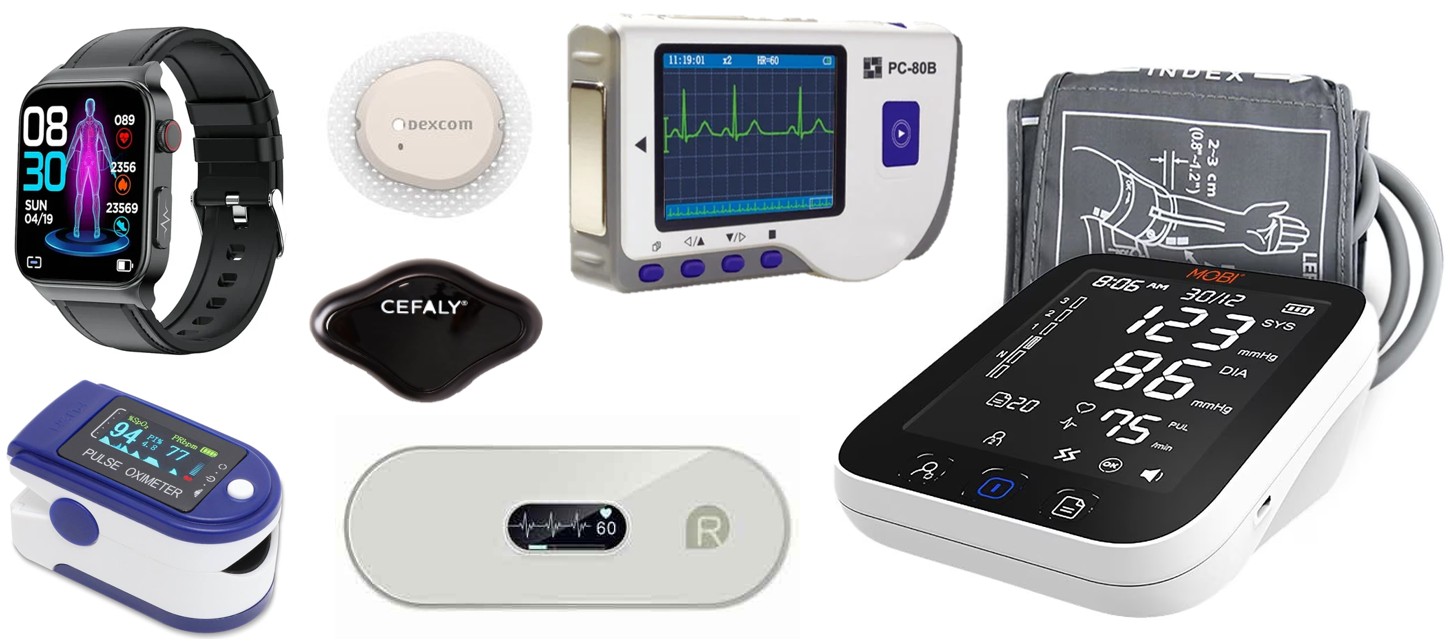}}
\caption{Proposed sample of wearable devices class.}
\label{fig3}
\end{figure}

\subsection{Wearable Devices}
Wearable devices are perhaps the most visible manifestation of IoT technology in healthcare~\cite{piwek2016rise}. These devices are designed to be worn on the body and continuously monitor physiological parameters. Examples include Fitness trackers, smartwatches, ECG, Pulse Oximeters, Blood Pressure, temperature, glucose, and neurological monitors, shown in Fig.~\ref{fig3}. Wearable devices contribute to preventive healthcare by enabling users to monitor their health metrics continuously. Studies have shown that these devices can improve patient engagement and adherence to treatment plans.

\begin{figure}[htbp]
\centerline{\includegraphics[width =\linewidth]{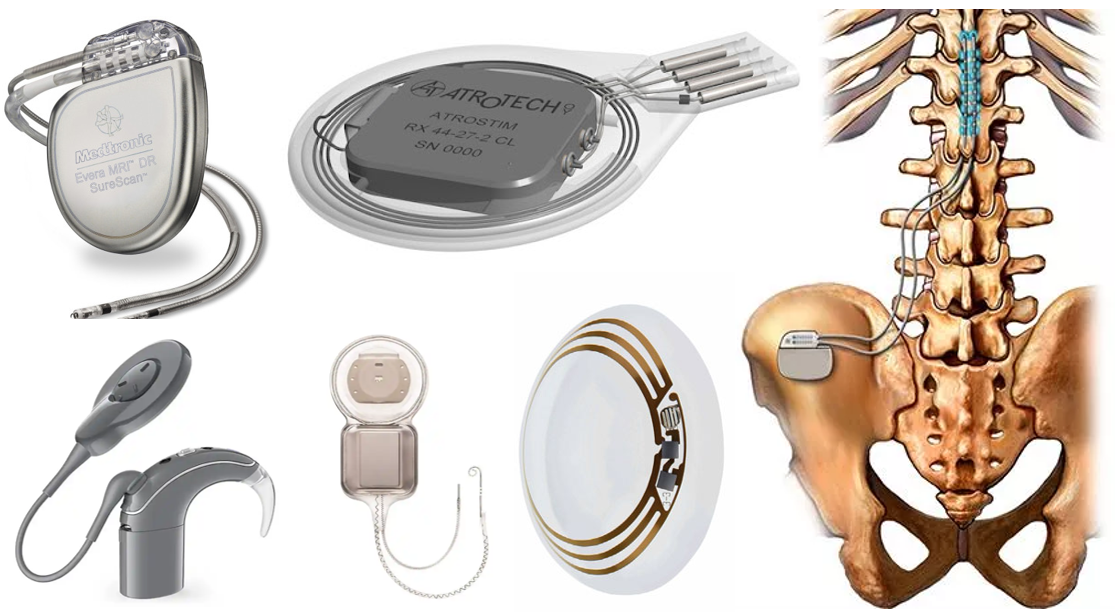}}
\caption{Proposed sample of Implanted devices class}
\label{fig4}
\end{figure}

\subsection{Implantable Devices}
Implantable devices are surgically placed inside the body to monitor and manage specific medical conditions~\cite{Casey2015-wd}. Examples include Cardiac Implantable Electronic Devices (CIEDs) and implantable cardioverter-defibrillators (ICDs), neurostimulators, spine cord simulators, cochlear implants, intraocular pressure sensors, and drug delivery systems, shown in Fig.~\ref{fig4}.
Implantable devices offer critical benefits for managing chronic conditions and can significantly enhance the quality of life for patients. They also enable remote monitoring, reducing the need for frequent hospital visits.

\begin{figure}[htbp]
\centerline{\includegraphics[width =\linewidth]{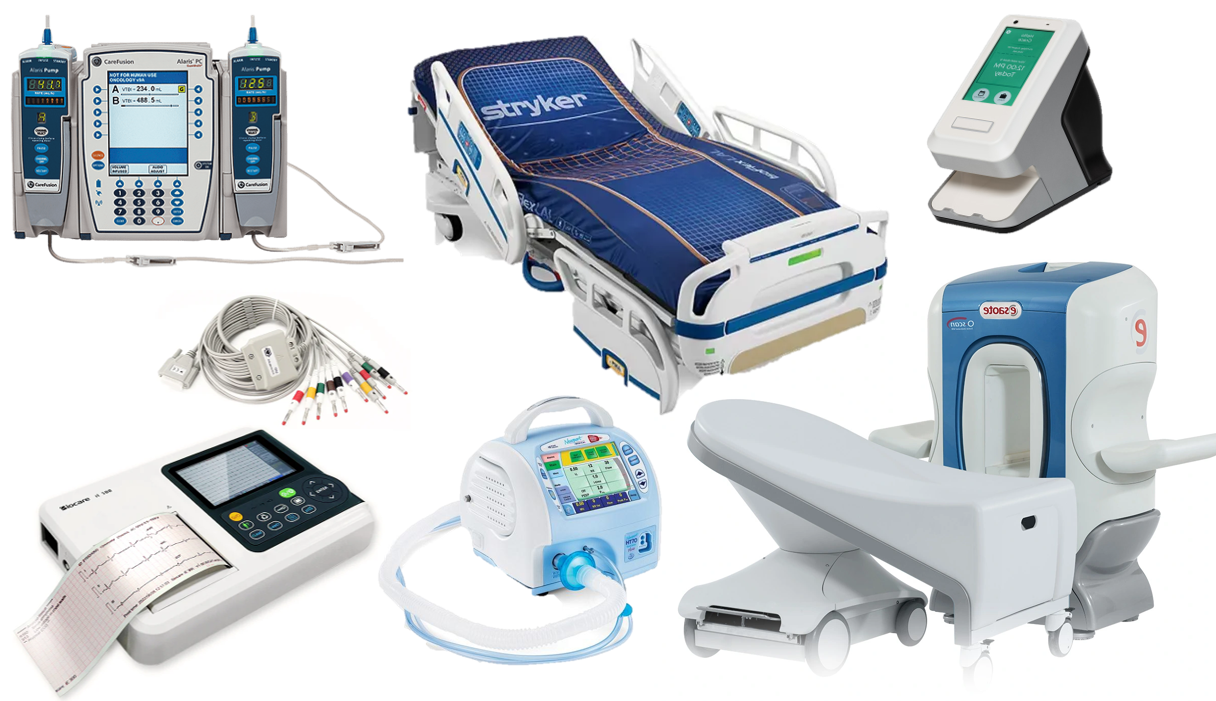}}
\caption{Proposed sample of smart devices class.}
\label{fig5}
\end{figure}

\subsection{Smart Devices}
Smart medical equipment encompasses various devices used in clinical settings for diagnostics, treatment, and patient care. Examples include smart infusion pumps, smart beds, automated dispensing systems, EKG machines, ventilators, MRI and CT scans that can share data in real-time, and monitors and displays. They can be programmed and monitored remotely, as shown in Fig.~\ref{fig5}.
Smart medical equipment improves the accuracy and efficiency of medical procedures, leading to better patient outcomes and streamlined clinical workflows~\cite{Dradrach2020-ws}.

\begin{figure}[htbp]
\centerline{\includegraphics[width =\linewidth]{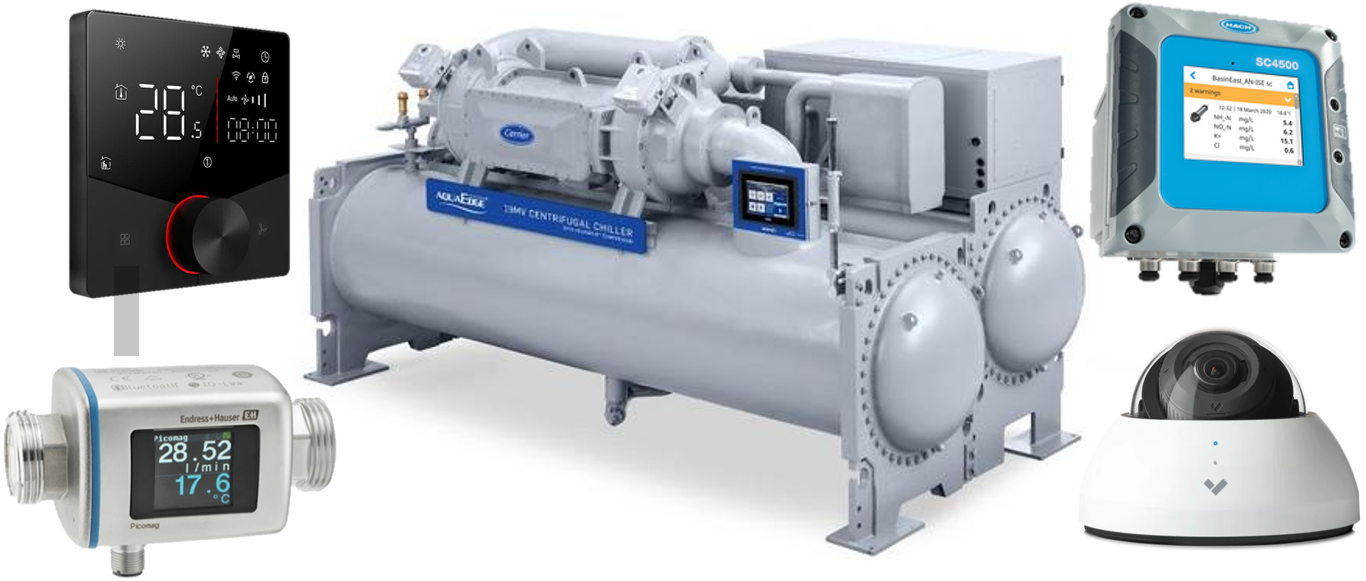}}
\caption{Proposed sample of ambient devices class.}
\label{fig6}
\end{figure}

\subsection{Ambient Devices}
Ambient devices consist of a wide range of equipment that supports the physical facility and the ambient environment of the healthcare building, such as thermostats, lights, CO2 monitors, water heaters, chillers, and HVAC. Additional remote monitoring devices, such as specialized cameras, microphones, and scales, are shown in Fig.~\ref{fig6}. 
Ambient medical devices have a very unstudied effect on the patient's health. However, they play a significant role in the accuracy of the other devices' readings and the associated medical diagnoses and interpretation~\cite{alshamrani2022iot}.

\begin{figure}[htbp]
\centerline{\includegraphics[width =\linewidth]{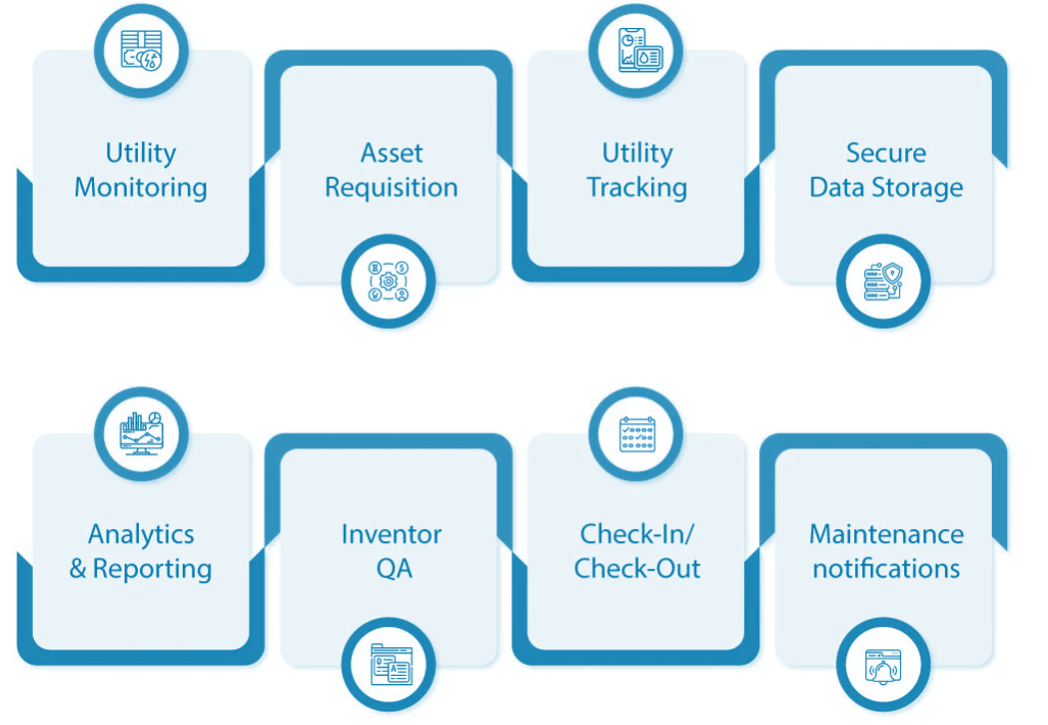}}
\caption{A sample of an IoT Based hospital assets tracking system by Rishabh\textregistered}
\label{fig7}
\end{figure}

\subsection{Operation Tools}
IoT devices also play a crucial role in optimizing hospital operations, enhancing efficiency, and ensuring patient safety. Such as Asset Tracking Systems, Real-Time Location Systems (RTLS),
Inventory Management, Hygiene Monitoring IoT Systems, shown in Fig.~\ref{fig7}~\cite{saradha2023study}~\cite{boyce2019challenges}

\subsection{Research and Development}
\begin{figure}[htbp]
\centerline{\includegraphics[width =\linewidth]{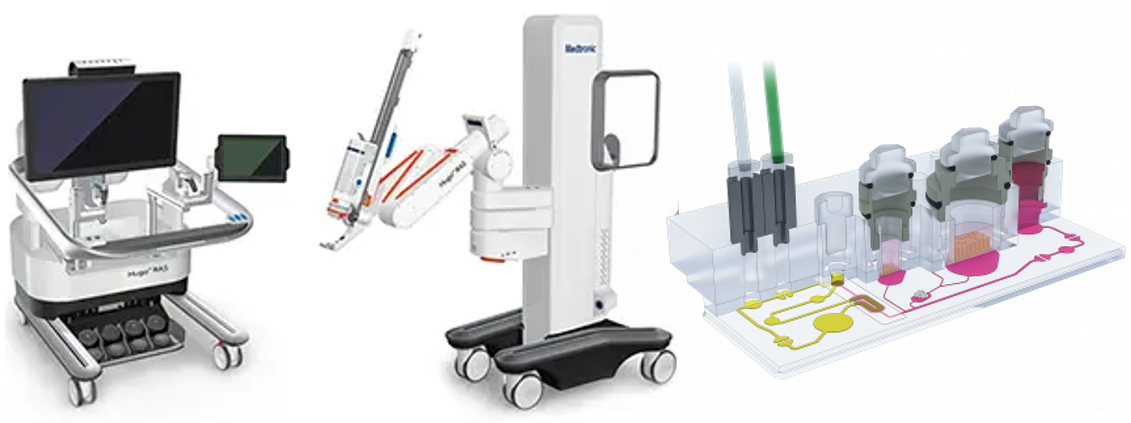}}
\caption{A sample of IoT Breseach and development tools in healthcare.}
\label{fig8}
\end{figure}

Research and Development class includes a wide range of devices used in clinical research for better treatment and diagnostics, Shown in Fig.~\ref{fig8}. Such as Organ-on-chip, Genetic Analyzers, and Robot-Assisted Surgery systems~\cite{bhuiyan2021internet}.

\section{Overview of Healthcare IoT Threats}

\subsection{Threats to: Wearable Devices}
The ubiquitous nature of wearable devices in healthcare makes them attractive targets for cyber attacks~\cite{dwivedi2023ten}. These devices often collect and transmit sensitive health data that, if compromised, can lead to significant privacy breaches and potential misuse of personal information. Common cybersecurity threats to wearable devices are:

\begin{itemize}
\item Data Interception: Since wearable devices often rely on wireless communication protocols such as Bluetooth, they are vulnerable to man-in-the-middle (MitM) attacks where an attacker can intercept and manipulate the data being transmitted between the device and the associated healthcare system.
\item Unauthorized Access: Many wearable devices lack robust authentication mechanisms, making it easier for unauthorized users to access the device and the data it holds.
\item Firmware Exploitation: Wearable devices require regular firmware updates to fix security vulnerabilities. However, if these updates are not properly secured, attackers can exploit them to inject malicious code or gain control over the device.
\item Data Leakage: Inadequate data encryption and poor security practices can result in unintentional data leakage, exposing sensitive health information to unauthorized parties.
\end{itemize}

\textbf{Mitigating} these threats require implementing strong encryption protocols, ensuring secure firmware update processes, and incorporating multi-factor authentication to protect against unauthorized access.

\subsection{Threats to Implantable Devices}
Integrating implantable devices in patient care brings significant benefits but exposes patients to unique cybersecurity threats. Given their critical functions, any compromise can have severe consequences. Common threats are:

\begin{itemize}
\item Device Hijacking: Attackers can gain unauthorized access to implantable devices, such as pacemakers or insulin pumps, potentially altering their settings to harm the patient.
\item Data Manipulation: By intercepting and modifying the data transmitted between implantable devices and healthcare systems, attackers can feed false information to healthcare providers, leading to incorrect diagnoses or treatments.
\item Denial of Service (DoS) Attacks: Implantable devices can be targeted with DoS attacks, which can turn off the device, potentially leading to life-threatening situations for patients reliant on continuous monitoring and treatment.
\item Firmware Exploitation: Like wearable devices, implantable devices require regular updates. If these updates are not securely implemented, they can be a vector for attacks where malicious code is introduced to the device.
\end{itemize}
\textbf{To mitigate} these risks, it is essential to adopt strong encryption protocols for data transmission, employ rigorous access control mechanisms, regularly update the firmware with secure methods, and conduct comprehensive security assessments of implantable devices~\cite{Smith2008-rf}.

\subsection{Threat to Medical Equipment and Logistic systems}
Smart medical equipment encompasses a wide range of stationary and mobile devices used in clinical settings for diagnostics, treatment, and patient care~\cite{Dradrach2020-hb}. Innovative medical equipment improves the accuracy and efficiency of medical procedures, leading to better patient outcomes and streamlined clinical workflows. Healthcare information systems improve care coordination and support data-driven decision-making, leading to more personalized and effective treatments~\cite{Yugi2016-pa}  IoT devices also play a crucial role in optimizing hospital operations, enhancing efficiency, and ensuring patient safety~\cite{DeSalvo2015-pv}, The common attacks to these systems are:
\begin{itemize}
\item Data Breaches: Unauthorized access to sensitive operational data can lead to data breaches, exposing information about hospital assets and operations.
\item Operational Disruptions: Cyberattacks on hospital operation tools can disrupt essential services, such as the availability of critical medical equipment and compliance with hygiene standards.
\item Ransomware: Hospital operation systems are also susceptible to ransomware attacks, where attackers encrypt operational data and demand a ransom for their release.
\end{itemize}

\textbf{Mitigating} these risks, hospitals should implement robust access controls, ensure regular security updates, and employ continuous monitoring of these IoT systems to detect and respond to security threats promptly.

Numerous studies indicate a rising trend in cyber attacks targeting healthcare IoT devices. For instance, a report by Check Point Research (2025) highlighted a $45\%$ increase in attacks on healthcare organizations compared to the previous year. The consequences of these attacks are severe, including compromised patient safety, financial losses, and damage to organizational reputation.

\section{Cybersecurity Threats to IoT Devices in Healthcare}
\begin{figure}[htbp]
\centerline{\includegraphics[width =\linewidth]{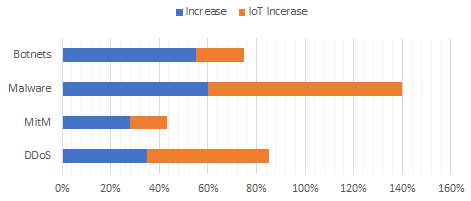}}
\caption{A sample statistics of the recent cyberattacks globally and specifically of Healthcare IoT systems and devices.}
\label{fig9}
\end{figure}

Recent statistics highlight the alarming frequency of such incidents. The 2024 Elastic Global Threat Report report by Cybersecurity Ventures revealed that the top attacks in this category are:
\begin{itemize}
\item Ransom Attacks.
\item Denial-of-Service (DoS) 
\item Distributed Denial of Service (DDoS).
\item Man-in-the-Middle Attack.
\item Malware and Botnets.
\end{itemize}
Where DDoS attacks targeting healthcare systems increased by $35\%$ compared to the previous year. Furthermore, insecure IoT devices contributed to over $50\%$ of these attacks, often leveraging outdated or unpatched firmware. According to the Healthcare Information and Management Systems Society (HIMSS), nearly $47\%$ of healthcare organizations reported experiencing a DDoS attack in the past two years, emphasizing the urgent need for enhanced network defenses. Furthermore, recent findings from Check Point Research (2022) showed a $28\%$ increase in MitM attacks targeting IoT-enabled medical devices. This trend is particularly concerning as such attacks can compromise real-time patient monitoring systems, potentially leading to life-threatening situations. The Ponemon Institute (2023) estimated that MitM attacks account for approximately $15\%$ of all reported security incidents in the healthcare sector.
Healthcare IoT malware incidents have increased in recent years. Data from Palo Alto Networks Unit 42 indicated a 60\% year-over-year increase in malware targeting IoT devices within healthcare environments in 2023. Botnets like Mirai and its variants have been adapted to exploit vulnerabilities in medical devices, highlighting the need for proactive threat detection systems. Additionally, a study by Symantec revealed that 83\% of healthcare IoT devices run on outdated operating systems, making them prime targets for malware exploitation, the overall state is shown in Fig.~\ref{fig9}.

\section{Solutions and Mitigation Strategies}
The healthcare sector's reliance on IoT devices necessitates robust and scalable solutions to counter evolving cybersecurity threats. This section delves into the strategies and technologies designed to address these challenges, focusing on enhancing device security, network resilience, and organizational readiness.

\subsection{Enhancing Device Security}
\begin{figure}[htbp]
\centerline{\includegraphics[width =\linewidth]{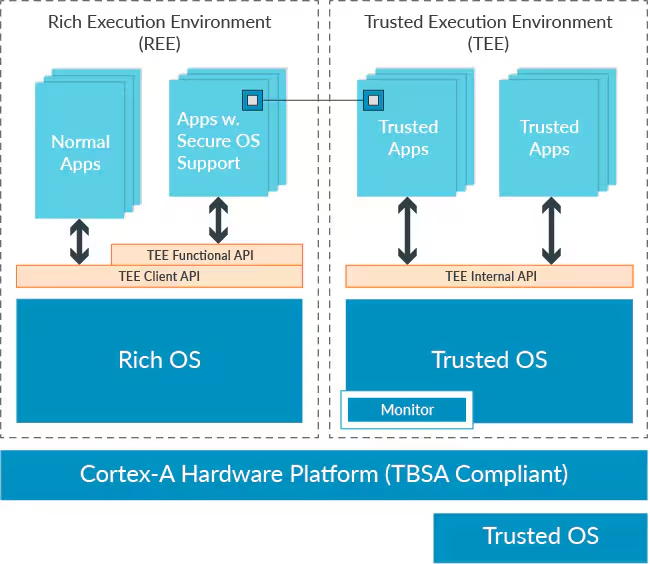}}
\caption{The Arm Trusted Zone layers Example Diagram.}
\label{fig10}
\end{figure}

Starting with simple devices, teams must prioritize security-by-design principles, incorporating features such as secure boot, firmware updates, and hardware-based encryption. Advanced tools like \textit{Arm TrustZone} and \textit{Intel SGX} can isolate critical applications and data, provide code monitoring, and prevent unauthorized access even if a device is compromised, as shown in Fig.~\ref{fig10}.

\subsection{Robust Authentication Mechanisms}
\begin{figure}[htbp]
\centerline{\includegraphics[width =\linewidth]{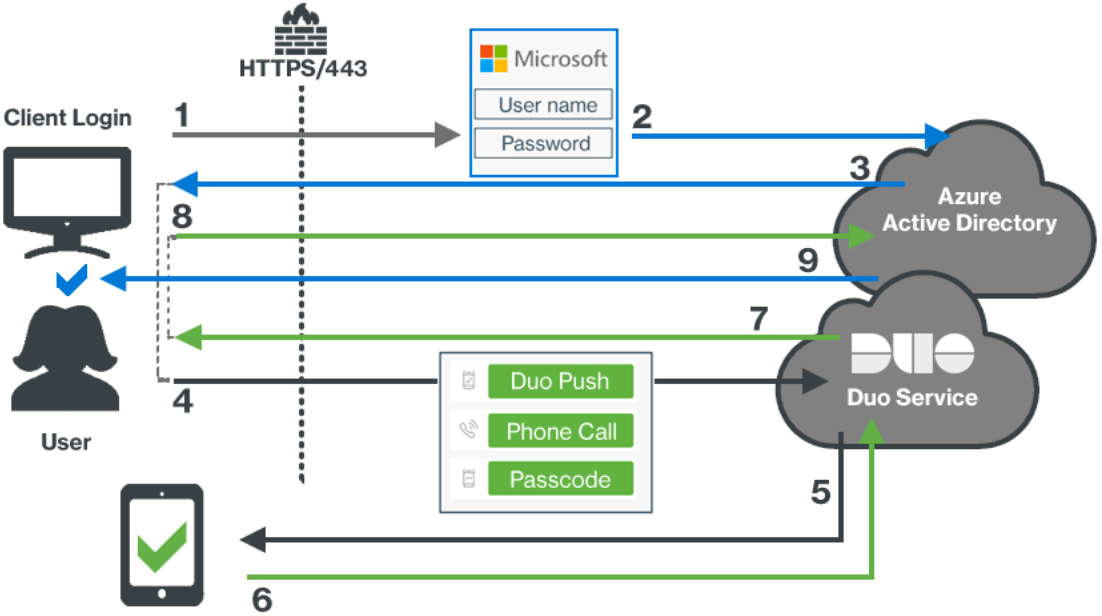}}
\caption{Authentication steps 1) user accesses services using Entra ID, 2) user submits primary credentials, 3) user selects the Duo EAM method, 4) user receives the Duo Prompt and submits factor selection, 5) user receives Duo Push authentication request on device, 6) authentication approval returned to Duo service, 7)secondary authentication approval returned to client, 8) client sends Duo approval back, and 9) Entra ID grants access.}
\label{fig11}
\end{figure}

Implementing multi-factor authentication (MFA) and biometric verification can reduce unauthorized access to IoT devices. Public Key Infrastructure (PKI) can also provide a scalable framework for authenticating devices and encrypting communications. Current Examples would be \textit{Ripple}, \textit{Okta}, \textit{Cisco Duo}, \textit{Microsoft Entra ID}, and \textit{IBM Verify}. An example of the nine steps of the authentication process of \textit{Cisco Duo} and \textit{Microsoft Entra ID} is shown in Fig.~\ref{fig11}.

\subsection{Network Security Enhancements}
Healthcare systems should deploy micro-segmentation to isolate IoT devices within virtual networks, minimizing the potential spread of malware. Intrusion detection systems (IDS) and intrusion prevention systems (IPS) that leverage artificial intelligence (AI), some classic tools such as \textit{Snort} or \textit{Zeek} that became standards or popular current tools such as \textit{AirMagnet}, \textit{GuardDuty}, \textit{Azure  IDPS}, \textit{Cisco NGIPS}, and \textit{NSFocu-NG} can provide real-time threat detection and response capabilities.

\subsection{Adopting Blockchain Technology}
\begin{figure}[htbp]
\centerline{\includegraphics[width =\linewidth]{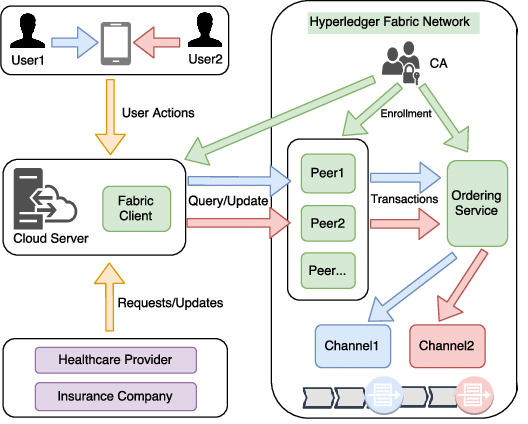}}
\caption{The Hyperledge Fabric system block diagram - Block Chain based solution for IoT devices.}
\label{fig14}
\end{figure}

Blockchain can provide decentralized and tamper-proof record-keeping for healthcare IoT devices. For example, solutions like \textit{Hyperledger Fabric}, which is an open-source enterprise-grade permissioned distributed ledger technology (DLT) platform, can ensure the integrity of medical data while supporting secure device-to-device communication. The system block diagram is shown in Fig.~\ref{fig14}.

\subsection{Regulatory and Compliance Frameworks}
\begin{figure}[htbp]
\centerline{\includegraphics[width =\linewidth]{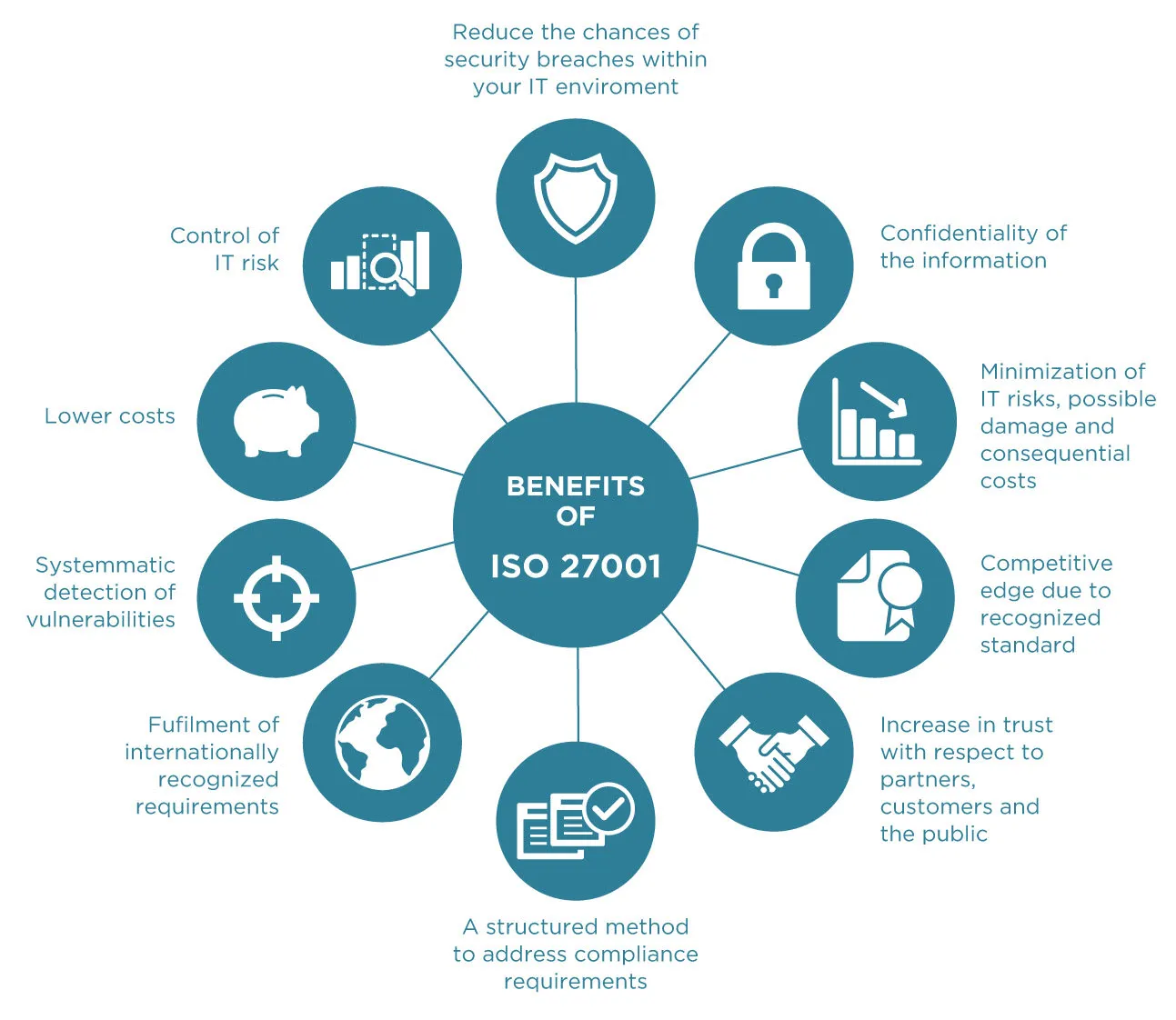}}
\caption{ISO/IEC 27001 Information Security Components.}
\label{fig15}
\end{figure}

Healthcare organizations must adhere to established cybersecurity frameworks like the \textit{NIST Cybersecurity Framework} and \textit{ISO/IEC 27001}~\cite{9528421} shown in Fig.~\ref{fig15}. These guidelines provide best practices for risk management, incident response, and data protection. Regular audits and compliance checks can further ensure adherence to these standards.

\subsection{User Awareness and Training}
Human errors often contribute to cybersecurity incidents. Comprehensive training programs for healthcare staff can mitigate risks by fostering awareness of phishing scams, social engineering, and safe device usage practices. Interactive platforms like \textit{KnowBe4}, \textit{Traliant}, and \textit{egress} can deliver tailored cybersecurity training to healthcare employees.

\subsection{Cutting-Edge Threat Detection Tools}
\begin{figure}[htbp]
\centerline{\includegraphics[width =\linewidth]{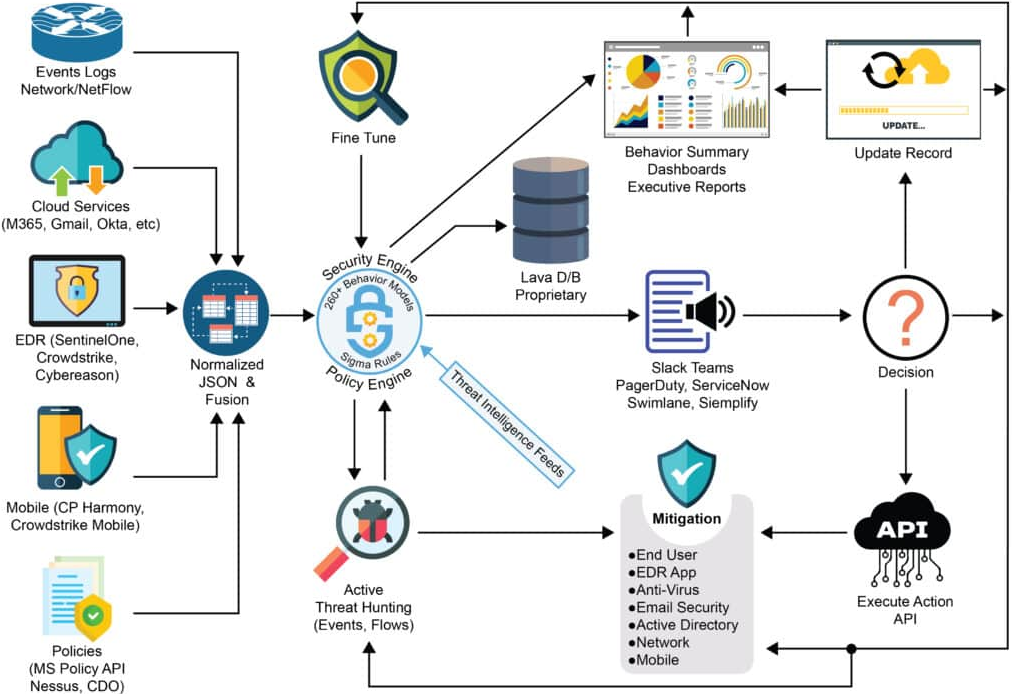}}
\caption{PTP’s SOAR implementation show case scenario.}
\label{fig16}
\end{figure}

Emerging technologies like Extended Detection and Response (XDR) and Security Orchestration, Automation, and Response (SOAR) platforms~\cite{Mazhar2023-sa} offer integrated and automated threat management and respond to security threats. Tools such as \textit{CrowdStrike Falcon} and \textit{Palo Alto Networks} can help healthcare organizations proactively identify and mitigate advanced threats. An Implementation example of SOAR-based security motoring  PTP-SOAR is shown in Fig.~\ref{fig16}.

\subsection{IoT-Specific Security Configurations}
\begin{figure}[htbp]
\centerline{\includegraphics[width =\linewidth]{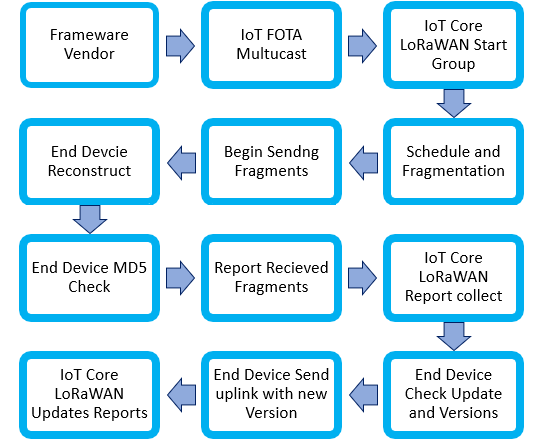}}
\caption{An example of and IoT FOTA based process block diagram of firmware update.}
\label{fig17}
\end{figure}

Custom configurations tailored to IoT devices can significantly enhance security. Techniques like lightweight encryption (e.g., elliptic-curve cryptography) ensure secure communication without burdening resource-constrained devices. Additionally, secure firmware over-the-air (FOTA) update mechanisms can promptly address vulnerabilities. FOTA technology allows manufacturers to upgrade and update device firmware wirelessly. FOTA-capable devices can download upgrades and updates directly from the service provider, usually taking 3–10 minutes, depending on the user’s connection speed and the file size. The process of  IoT Core for LoRaWAN performs the FOTA process for IoT end devices, shown in Fig.~\ref{fig17}.

\section{Future Research Directions}

While significant strides have been made in addressing cybersecurity challenges in IoT-enabled healthcare systems, there are pressing needs for advanced research and development. Future research directions include:
\subsection{Machine Learning and AI-Driven Threat Detection}
Developing more sophisticated machine learning models for detecting and mitigating cyber threats in IoT ecosystems is critical. Techniques such as deep learning and federated learning can analyze vast datasets to identify anomalies in real time. Moreover, AI-driven tools can adapt to new attack vectors, offering proactive security measures rather than reactive responses. Integrating natural language processing (NLP) to detect phishing or social engineering attempts could also enhance IoT device resilience.

\subsection{Lightweight Cryptographic Algorithms}
Given the resource constraints of IoT devices, lightweight cryptographic algorithms need further exploration. Solutions such as Elliptic Curve Cryptography (ECC) or lattice-based cryptography are promising for ensuring secure communication while maintaining low computational overhead.

\subsection{Quantum-Resilient Security Protocols}
As quantum computing becomes a reality, many existing encryption methods will be obsolete. Research into post-quantum cryptography is essential for preparing IoT devices in healthcare for future threats. Exploring hybrid cryptographic models that combine classical and quantum-resistant techniques could provide an interim solution.

\subsection{Secure Interoperability Frameworks}
Healthcare IoT devices often operate in heterogeneous environments, requiring seamless, secure communication. Developing frameworks that ensure interoperability without compromising security is critical. To achieve this goal, standards such as the Fast Healthcare Interoperability Resources (FHIR) can be extended with enhanced security features.

\subsection{ Behavioral Analytics and Insider Threat Mitigation}
Research into behavioral analytics can offer insights into insider threats, which remain a significant challenge. Advanced machine learning models can analyze user and device behavior to identify anomalies, malicious activity, or insider threats.

\subsection{Energy-Efficient Security Mechanisms}
IoT devices in healthcare often operate under strict power constraints. Research into energy-efficient security mechanisms, such as adaptive encryption algorithms and context-aware security protocols, is essential to extend device longevity while maintaining robust protection.

\section{Conclusion}
IoT devices have revolutionized the healthcare industry, providing unprecedented opportunities for real-time monitoring, patient care, and operational efficiency. However, this transformation is accompanied by significant cybersecurity challenges that demand urgent attention. 
Adopting a multi-layered security approach incorporating regulatory compliance, user awareness, and cutting-edge technologies is essential to mitigate these risks effectively. Furthermore, fostering collaboration between manufacturers, healthcare providers, and cybersecurity experts can lead to the development of resilient systems that safeguard sensitive patient data and ensure uninterrupted healthcare services.
Future research must prioritize exploring machine learning and AI-based threat detection systems, lightweight cryptographic solutions, and quantum-resilient protocols. Additionally, integrating decentralized security architectures and enhancing interoperability frameworks can pave the way for a more secure and efficient IoT-enabled healthcare ecosystem. By addressing these challenges, the healthcare industry can harness the full potential of IoT while maintaining the highest standards.

\section*{Acknowledgment}
Special thanks to S. Bay from Johnson Controls for the data, technical reports, and case studies. Thanks and appreciation to the Applied Machine Learning and Intelligence lab in the School of Information for the comprehensive review and validation tasks.

\bibliographystyle{ieeetr}
\bibliography{references}

\end{document}